\documentclass[10pt, conference]{IEEEtran}
\pdfoutput=1
\IEEEoverridecommandlockouts
\usepackage{cite}
\usepackage{amsmath,amssymb,amsfonts}
\usepackage{algorithmic}
\usepackage{graphicx}
\usepackage{textcomp}
\usepackage{xcolor}
\usepackage{multirow}
\usepackage{pifont}
\usepackage{algorithm}
\usepackage{algorithmic}
\usepackage{tikz}
\usepackage{pgfplots}
\usepackage{cleveref}
\usepackage{float}
\usepackage{placeins}
\usepackage[shortcuts]{extdash}
\usepackage{url}
\usepackage{booktabs}
\usetikzlibrary{arrows.meta, positioning, shapes, backgrounds, decorations.pathmorphing}
\definecolor{pastelgreen}{rgb}{0.47, 0.87, 0.47}
\definecolor{pastelred}{rgb}{0.98, 0.5, 0.45}
\def\BibTeX{{\rm B\kern-.05em{\sc i\kern-.025em b}\kern-.08em
    T\kern-.1667em\lower.7ex\hbox{E}\kern-.125emX}}
\begin{document}

\title{Bursts and Triggers: Socially-Driven Activity in Open-Source Co-Editing Networks\\
}

\author{
\IEEEauthorblockN{Lisi Qarkaxhija\textsuperscript{1}, Maximilian Capraro\textsuperscript{2}, Stefan Menzel\textsuperscript{3}, Bernhard Sendhoff\textsuperscript{3}, Ingo Scholtes\textsuperscript{1}}
\IEEEauthorblockA{
\textsuperscript{1}CAIDAS, Julius-Maximilians-Universit\"at W\"urzburg, Germany\\
\textsuperscript{2}DATEV eG, N\"urnberg, Germany\\
\textsuperscript{3}Honda Research Institute Europe, Offenbach am Main, Germany\\
\{lisi.qarkaxhija, ingo.scholtes\}@uni-wuerzburg.de, maximilian.capraro@datev.de,\\
\{stefan.menzel, bernhard.sendhoff\}@honda-ri.de}
}

\maketitle

\begin{abstract}
The long-term sustainability of Open Source Software (OSS) communities depends on the activity of their developers, yet the social mechanisms driving this collective behavior remain poorly understood. Analyzing commit histories across 51 major OSS communities, we find that developer contributions are strongly ``bursty'' in time. To test whether this burstiness reflects social responsiveness rather than individual habit alone, we model developer interactions as temporal co-editing networks and introduce a method to detect activity triggers, episodes in which one developer editing another's code is followed by an unusually rapid response, and the cascades they form. Benchmarking these against a null model that destroys the temporal ordering of co-edits while preserving each developer's activity rate, we find statistically significant cascades in 28 of 51 projects (55\%) under our default configuration, though prevalence ranges from 24\% to 82\% across detection thresholds. Whether a project exhibits significant cascades is governed primarily by its scale rather than governance or commit concentration. Finally, as a secondary application, we test whether these signals inform developer churn: features capturing the recent (in)activity of a developer's collaborators add some predictive value, but a developer's own inactivity dominates, propagation over the co-editing graph adds little, and simple models match graph neural networks. Our results characterize developer responsiveness as a measurable component of collective OSS dynamics.
\end{abstract}


\section{Introduction}
\label{sec:intro}

Open source software (OSS) development represents one of the most successful forms of large-scale collaborative knowledge work in the digital age~\cite{raymond2001cathedral, weber2004success}. 
Understanding the social mechanisms that enable thousands of distributed developers to effectively coordinate their work is fundamental to comprehend how complex software systems emerge from decentralized collaboration~\cite{Crowston_Howison_2005, von2003community}. 
To this end, numerous studies in empirical software engineering have considered human and social aspects such as the motivation of individual developers~\cite{lakhani2003hackers, hertel2003motivation}, the influence of project governance structures and team size~\cite{o2003guarding,shaikh2017governing,scholtes2016aristotle} as well as the role of (temporal) collaboration and communication networks~\cite{bird2006mining, gote2019git2net}.

An interesting yet less studied aspect of collective social dynamics in OSS communities are temporal patterns of developer contributions, i.e. at which times developers contribute to a project.
Human activities are often characterized by ``bursty'' dynamics, where short periods of intense activity are followed by long periods of inactivity \cite{goh2008burstiness,karsai2018bursty}. 
Such patterns are a hallmark for the influence of complex memory and triggering effects,  i.e. one action influencing the timing of subsequent actions,  which invalidates simple Poissonian models for memoryless events independently occurring at a constant rate.
Such memory effects can naturally arise due to \emph{individual human behavior}, since humans are likely to exhibit bursts in which they perform sequences of similar actions within a short time period.
However, apart from this exogenous explanation, the same patterns can also be endogenously explained by \emph{collective human behavior} in social networks, where the action of one actor triggers subsequent events by its neighbors.
The importance of such social mechanisms to understand collective behavior has been highlighted in various contexts, including E-Mail and letter correspondence \cite{malmgren2009universality}, collective attention in social media \cite{de2020unraveling}.

\begin{figure}[htbp]
    \includegraphics[width=\linewidth]{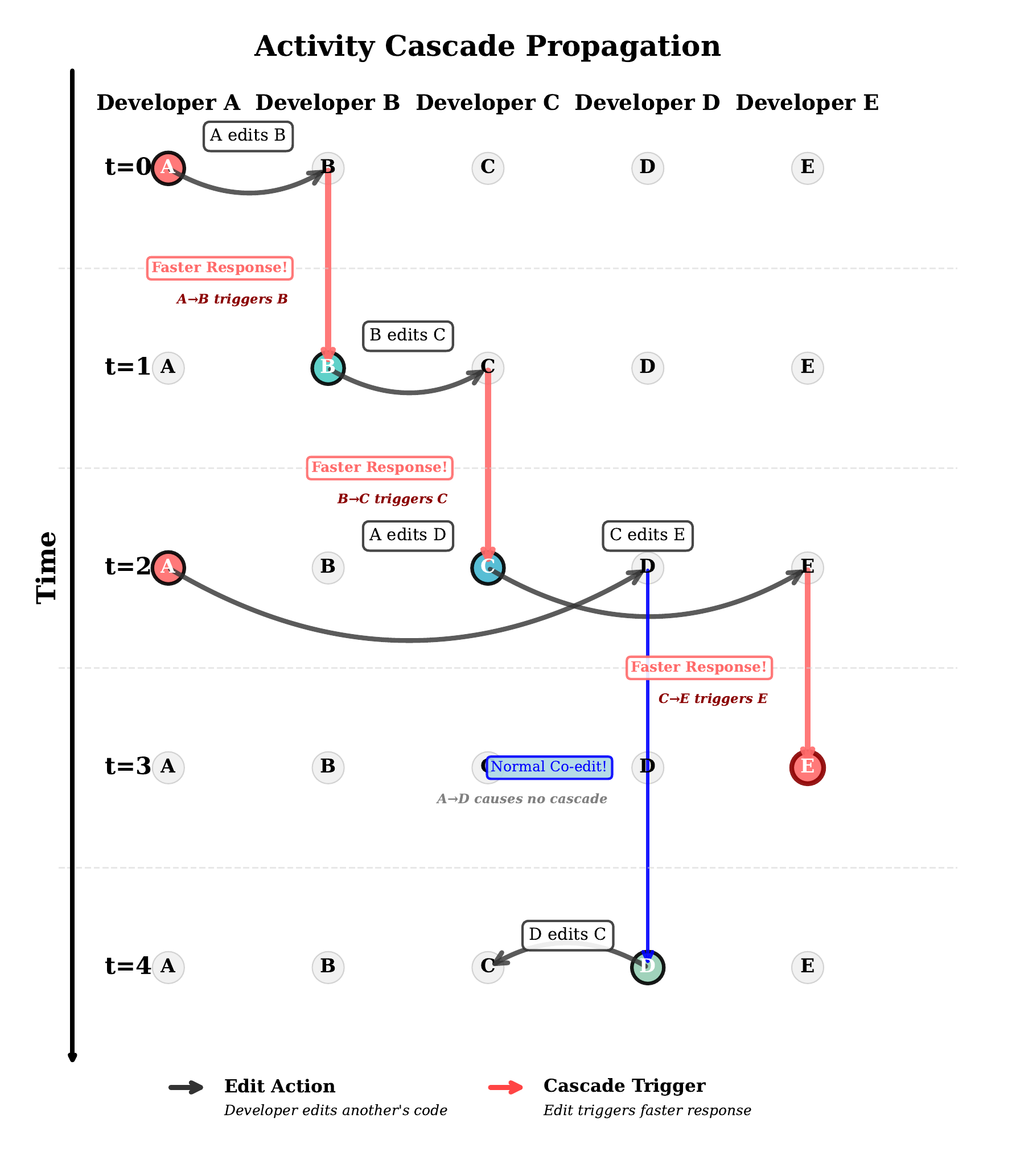}
    \caption{Visualization of activity cascades in software development. 
    Each row represents a time step, and each column a developer. 
    Arrows indicate co-edited code between developers, while vertical arrows show subsequent edits by the same developer. 
    A cascade is triggered when a developer ``responds'' faster than usual after his or her code is edited (highlighted with 'Faster Response!' and a trigger label). Co-edits that do not result in a faster response (e.g., A $\rightarrow$ D) are also shown, illustrating that not all edits necessarily trigger a cascade.}
    \label{fig:infographics}
\end{figure}

The presence of such collective dynamics in other types of human behavioral data raises the question whether similar forms of social influence drive temporal activity patterns of developers in OSS communities. 
Moreover, if such a collective phenomenon exists, it is interesting to study which type of interactions between developers mediate the underlying social influence, and whether it can lead to \emph{activity cascades} that propagate through the time-evolving collaboration network of OSS communities (see Figure~\ref{fig:infographics}).
Understanding the mechanisms behind such activity cascades could provide insights into how distributed teams self-organize and how individual actions influence collective behavior.
Moreover, discovering social mechanisms that trigger developer activity can shed light on the question why some developers leave a project, possibly informing the development of methods to predict developer churn or retention.
To the best of our knowledge these questions have not been studied in the context of collaborative software engineering, which constitutes the research gap that is addressed by our work.

\paragraph{Research Questions and Hypotheses} Closing this gap, our work is guided by three primary research questions:

\textbf{RQ1: Do commit activities of developers in OSS communities exhibit bursty patterns?} We hypothesize that the inter-event time distributions of commits in OSS projects exhibit a bursty pattern, indicating that developer contributions are not independent events. The presence of such a pattern, and thus the deviation from a Poissonian distribution,  indicates the presence of memory (i.e. the action of one developer influencing future actions of the \emph{same} developer) and/or social influence (i.e. the action of one developer influencing future actions of other developers).

\textbf{RQ2: Do activity cascades propagate along co-editing networks?} Beyond memory effects, we hypothesize that a possible explanation for the presence of bursty activity patterns is social influence that is mediated via co-editing networks. 
This assumes a social mechanism where developers are likely to exhibit a higher level of activity shortly after their code has been edited by other developers.
We further investigate whether this social influence transitively expands into activity cascades that propagate through the co-editing network.

\textbf{RQ3: Can the presence of activity cascades inform developer churn prediction?} 
We finally hypothesize that the social mechanism behind developer activity patterns studied in {\bf RQ2} can enhance our ability to forecast which developers are likely to leave a project.
This is based on the assumption that a lack of activity in the neighborhood of a given developer, and the resulting lack of social influence triggering future contributions, can ultimatively make it more likely that the developer leaves the project.

\paragraph{Research Contribution and Methodology} Addressing the questions outlined above, our work makes the following contributions:
\begin{itemize}
    \item Using a corpus of 51 major OSS communities covering at least four years, we analyze temporal commit patterns of developers. Our results show that inter-commit times exhibit bursty patterns both at the level of projects and individual developers.
    \item Motivated by this finding, we develop a method to detect activity cascades that traces the propagation of accelerated responses through the co-editing network. 
    \item Through an analysis of 51 diverse OSS projects, we show that activity cascades are a statistically significant phenomenon in 55\% of the analyzed projects. This suggests developers are highly responsive to code modifications, indicating implicit coordination that enables rapid collaborative responses in distributed environments.
    \item We finally show that insights from cascade analysis can enhance the prediction of developer churn, demonstrating the practical utility of our framework.
\end{itemize}

The remainder of this paper is structured as follows: \Cref{sec:related} reviews related work. \Cref{sec:methodology} presents our methods to analyze bursty commit patterns and detect activity cascades. 
In \Cref{sec:results} we present empirical results. \Cref{sec:churn} applies our insights to predict developer churn. 
In \cref{sec:discussion} we discuss implications and limitations of our work, before concluding in \cref{sec:conclusion}.

\section{Related Work}
\label{sec:related}

Our work builds upon several interconnected research streams that examine temporal collaboration patterns, social networks, and emergent coordination in OSS communities.

\paragraph{Collaboration Patterns and Temporal Networks in Software Projects}

Social network analysis has revealed how collaboration structure influences project outcomes in OSS.
\cite{zoller2020topology} show that the hierarchical group structure of collaboration networks influences long-term sustainability of Open Source communities, while \cite{scholtes2016aristotle} studied the interplay between the densification of co-editing networks and developer productivity.
\cite{Crowston_Howison_2005} find that the centralization of communication distinguishes projects, with larger communities tending toward more decentralized patterns.

The temporal dimension of these networks has gained increasing attention.
\cite{gote2019git2net} developed a method to mine time-stamped co-editing networks from Git repositories, with follow-up work showing that co-editing structure influences the timing of commits~\cite{gote2022big}.
\cite{ma2014dynamics} reveal power law distributions of inter-commit times, and \cite{sornette2014much} find heavy-tailed developer contributions, both indicative of collective dynamics.
Early contribution patterns are also strong predictors of sustained engagement~\cite{xiao2023early}, suggesting that the timing of interactions may be as important as their content.

\paragraph{Cascade Analysis in Complex Networks}
While studying the structure of temporal collaboration networks is important, a crucial and less investigated further aspect is that those networks are the substrate of dynamical processes that operate on them.
An important class of processes are \emph{cascades} by which information, failures, diseases or social behavior propagates through complex networks.
This phenomenon has been extensively studied in various complex networks, including social media~\cite{10.1145/2229012.2229058, doi:10.1126/science.aap9559, gomez2012inferring}, financial markets~\cite{10.1257/aer.104.10.3115,10.1257/jel.20151228}, and biological networks~\cite{pastor2015epidemic}. 
Traditional cascade analysis focuses on information propagation~\cite{kempe2003maximizing} or behavioral contagion~\cite{centola2010spread}, where actions by one agent influence the probability of similar actions by connected agents.
However, fewer studies have analyzed the propagation of cascades in the context of collaborative software projects.
Being a natural application of models of failure cascades in technical systems, some studies have considered the propagation of changes or failures in software dependency networks.
\cite{Geipel2009} studied cascades of changes propagating through the dependency networks of Open Source projects.
More recently, \cite{shehata2025cascading} investigated how project failures propagate across the dependency network of the Maven Central ecosystem.
Neither of the two works did link this to the developer collaboration networks of the underlying projects.
While \cite{sornette2014much} studied a ``cascading model of productive activity'' in terms of a Hawkes Poisson process, this model deviates from previously cited works on cascades in the sense that it did not consider the propagation of cascades in a (social) network.
Recently, \cite{schueller2024modeling} demonstrated how risks propagate through dependency and maintainer networks, showing that social factors can amplify technical vulnerabilities.

Our cascade detection method differs from established cascade models in network science.
Threshold models~\cite{watts2002simple} define cascades as binary adoption decisions propagating through a static network when a critical fraction of neighbors has adopted; informational cascade models~\cite{bikhchandani1992theory} study sequential rational agents who ignore private signals in favor of observed herd behavior.
Both assume discrete decision-making, whereas our setting involves continuous activity where developers do not make explicit adopt/reject choices.
Self-exciting point processes (Hawkes processes)~\cite{hawkes1971spectra} capture temporal clustering of events via parametric triggering kernels but do not model propagation through a network.
Our method combines network structure with temporal analysis: we trace cascades along co-editing edges using a nonparametric, per-developer baseline (percentile rank against each developer's own inter-commit distribution), requiring no distributional assumptions.

\paragraph{Modeling and Predicting Developer Turnover}

Developer turnover has been studied through multiple lenses.
\cite{betti2025dynamics} show that leadership changes are common in OSS projects and significantly impact success metrics, providing context for how individual departures can trigger broader changes in project dynamics.
Value-based discussions in OSS projects can predict changes in contributor turnover patterns~\cite{jamieson2024predicting}, highlighting the importance of social dynamics beyond technical considerations.
Survival analysis has provided insights into retention factors~\cite{lin2017developer}, and more recently \cite{chang2024condygnn} use graph neural networks to predict developer departures.
These approaches focus on predicting turnover from individual or structural features; our work complements them by examining how collective social dynamics may influence retention.

\paragraph{Research Gap}
While existing research has established foundations for understanding social dynamics in software projects, most studies focus on network properties or individual-level factors, with limited attention to dynamic social processes that emerge from direct developer interactions.
To the best of our knowledge, no studies have examined how developer activity propagates through temporal collaboration networks and how this process shapes collective contribution dynamics, nor has this been considered as a factor influencing developer turnover.
Addressing these gaps, our work introduces cascade analysis as a tool to understand collective social dynamics in software development, positioning social dynamics as an endogenous factor of collaborative behavior that can be measured and analyzed.

\section{Methodology}
\label{sec:methodology}

Our methodology is twofold: 
First, we analyze the temporal dynamics of commit activity to test for burstiness, which informs our subsequent investigation. 
Second, based on these findings, we develop a framework to detect and analyze activity cascades that propagate within co-editing networks.

\subsection{Burstiness Analysis (RQ1)}
To investigate RQ1, we analyze temporal patterns of commits to determine if they exhibit burstiness.
Burstiness refers to the phenomenon where a large number of events occur clustered within short time periods, separated by long periods of inactivity. 
We quantify this using the burstiness coefficient, $B$, proposed by~\cite{goh2008burstiness}.
Given a sequence of inter-event times, $\tau = \{\tau_1, \tau_2, ..., \tau_n\}$ capturing time differences between consecutive commits, the burstiness coefficient is given as $B = (\sigma_{\tau} - \mu_{\tau}) / (\sigma_{\tau} + \mu_{\tau})$, where $\mu_{\tau}$ and $\sigma_{\tau}$ are the mean and standard deviation of inter-event times, respectively.
The value of $B$ ranges from -1 to 1: For $B \approx 1$ the process is highly bursty, for $B \approx 0$ events occur independently (following a Poissonian distribution), and for $B \approx -1$ the process is highly regular or periodic.

We calculate $B$ for the inter-event times of commits at two levels:
\begin{enumerate}
    \item \textbf{Project-Level Burstiness:} We treat all commits in a project as a single time series, independent of commit authors.
    \item \textbf{Individual-Level Burstiness:} For each developer, we calculate $B$ based on their personal commit history, i.e. we consider inter-event times for commits of the same developer.
\end{enumerate}

To validate that individual-level burstiness  constitutes a pattern, we compare the observed coefficient against a null model in which we randomly shuffle the time stamps of all commits within a project.
By this, we selectively destroy individual-level burstiness but preserve other aspects such as the total commit activity, project-level inter-event times, as well as the number of commits per developer.

\subsection{Co-editing Network Construction}

Our analysis of activity cascades (RQ2) relies on co-editing networks constructed from Git commit histories. 
A directed temporal edge $A \to B$ is created at time $t$ if developer $A$ commits a change to a line of code previously edited by developer $B$. 
We use git2net~\cite{gote2019git2net} to extract these file-level co-editing relationships from the default branch of each repository. Following~\cite{Gote2021thesis}, merge commits that modify more than 1,000 files are excluded as they primarily combine existing work rather than introduce new contributions; all remaining commits, including standard merges, are retained.
The temporal nature of edges is crucial, as it allows us to trace the sequential flow of interactions and detect accelerated responses as described below.

\subsection{Trigger Event and Cascade Detection (RQ2)}
Bursty activity patterns can possibly be explained by a triggering mechanism by which the commit of one developer triggers subsequent commits of other developers.
Our cascade detection method is designed to identify such triggering events and trace their propagation.
The idea is that a cascade is initiated when a co-editing event prompts an unusually rapid response from the edited developer.

\begin{algorithm}
\caption{Trigger Event Detection}
\label{alg:trigger_detection}
\begin{algorithmic}[1]
\REQUIRE Developer commit histories $H$, co-editing event $e = (editor, edited, edit\_time)$
\ENSURE Boolean indicating if $e$ is a trigger event
\STATE $C \leftarrow H[edited]$ \COMMENT{All commit times of the edited developer}
\STATE $B \leftarrow \{t \in C : t < edit\_time\}$ \COMMENT{Commits before the edit}
\STATE $A \leftarrow \{t \in C : t > edit\_time\}$ \COMMENT{Commits after the edit}
\IF{$|B| \geq 2$ \AND $|A| \geq 1$}
    \STATE $I \leftarrow$ intervals between consecutive times in $B$
    \STATE $r \leftarrow$ percentile rank of $(A[0] - edit\_time)$ among $I$
    \IF{$r \leq 25$}
        \STATE \textbf{return true} \COMMENT{Unusually fast response}
    \ENDIF
\ENDIF
\STATE \textbf{return false}
\end{algorithmic}
\end{algorithm}

\Cref{alg:trigger_detection} details our method for identifying trigger events. 
When developer A edits B's code, we examine B's response time relative to B's historical commit patterns.
We compute the percentile rank of the response interval (time to B's next commit) compared to B's typical inter-commit intervals.
A response in the lowest quartile (rank $\leq 25\%$) is classified as a trigger.
We validate this threshold choice through sensitivity analysis: varying the percentile from 10th to 50th and the developer filter from 10\% to 30\%, we find that the proportion of projects with significant cascades ranges from 24\% to 82\%, with the qualitative finding,  that a substantial subset of projects exhibits significant cascades,  holding across all 9 parameter combinations (see Section~\ref{sec:results}).

\begin{algorithm}
\caption{Cascade Chain Detection}
\label{alg:cascade_detection}
\begin{algorithmic}[1]
\REQUIRE Commit histories $H$, co-editing events $E$, top developers $T$
\ENSURE Set of cascade chains $C$
\STATE $C \leftarrow \emptyset$
\STATE $S \leftarrow$ events in $E$ sorted by time
\FOR{each $e$ in $S$}
    \IF{$e.editor \notin T$ or not \textsc{IsTrigger}($H, e$)}
        \STATE \textbf{continue}
    \ENDIF
    \STATE $chain \leftarrow [e]$
    \STATE $cur \leftarrow e$
    \WHILE{True}
        \STATE $n \leftarrow$ first event in $S$ with $editor = cur.edited$ and $edit\_time > cur.edit\_time$
        \IF{$n$ is None or not \textsc{IsTrigger}($H, n$)}
            \STATE \textbf{break}
        \ENDIF
        \STATE append $n$ to $chain$; $cur \leftarrow n$
    \ENDWHILE
    \IF{$|chain| > 1$}
        \STATE add $chain$ to $C$
    \ENDIF
\ENDFOR
\STATE \textbf{return} $C$
\end{algorithmic}
\end{algorithm}

We use \cref{alg:cascade_detection} to trace triggers initiated by the top developers selected above and identify activity cascades.
An activity cascade is a sequence of trigger events: 
A edits B, B responds quickly (trigger 1); then B edits C, and C responds quickly (trigger 2), and so on (see Figure~\ref{fig:infographics}). 
We only consider cascades of depth greater than one, i.e. a sequence of at least two trigger events.

\subsection{Statistical Validation Framework}

To establish the statistical significance of observed cascade patterns, we use permutation testing, a non-parametric approach that avoids distributional assumptions.

We consider a \textbf{Temporal Shuffling} null model, which randomly permutes the timestamps of co-editing links while preserving the developers involved in a co-editing relationships. 
With this we test if cascades depend on the temporal ordering of co-edits.
We repeat the detection of triggering and cascade events for shuffled commits. 
For each project we compare empirical cascade counts against the distribution of cascades detected in 100 iterations of the null model.
We calculate empirical $p$-values by determining the fraction of shuffled instances where expected cascade counts equal or exceed empirically observed count.

Beyond statistical significance, in statistical analysis it is crucial to additional consider effect size to assess the importance of a finding.
To this end, we calculate Cohen's $d$, which quantifies the difference between observed cascade count and the expected cascade count null model means in terms of standard deviations.

To validate our cascade detection framework, we test it in two synthetically generated networks with known ground truth patterns. 
This evaluation demonstrates the method's accuracy, sensitivity to specific cascade types, and robustness against noise.

Our synthetic validation employs two synthetically generated networks: (1) a \textbf{random network} serving as a negative control, where developers have random commit histories and co-editing events occur between random developer pairs at random times, designed to confirm our method does not produce false positives; and (2) a \textbf{cascade network} engineered to test the detection of cascades, where developers have characteristic commit intervals creating regular development rhythms, and cascades are planted by inserting co-editing events immediately before expected commit times to create unusually short response intervals \footnote{Code and data to reproduce our experiments can be accessed via following link: \url{https://doi.org/10.5281/zenodo.21073503}.}.

\subsection{Dataset and Preprocessing}

Our repositories are drawn from the curated corpus of \cite{gote2022big}, who developed a multi-stage pipeline to select collaborative software development projects from GitHub. Starting from over 125 million repositories in the GHTorrent database~\cite{gousios2013ghtorent}, the pipeline applies the filtering criteria of \cite{kalliamvakou2014promises} to exclude personal, inactive (fewer than 50 commits or a span shorter than 100 days), and very small (single-developer) repositories. Forked repositories are removed to avoid analyzing duplicate commit histories. The resulting set of approximately 1.8 million original, collaborative projects is further filtered by programming language to retain only software development projects, yielding a corpus of 201 repositories stratified by team size to cover the full range of project scales on GitHub.

From this corpus, we selected all repositories satisfying two additional criteria: a minimum of 4 years of development history and at least 50 unique contributors, both necessary for meaningful temporal cascade analysis. This yielded our final dataset of 51 repositories spanning diverse domains including machine learning frameworks, system software, web development tools, and developer utilities (see Table~\ref{tab:repo_stats}).

For each repository, we extract complete commit histories and apply author disambiguation using the method of \cite{gote2021gambit} to handle developers contributing under multiple identities, preventing artificial fragmentation of activity patterns.

\paragraph{Top Developer Selection}
Following established evidence for the Pareto principle in OSS activity~\cite{goeminne2011evidence}, we focus our cascade detection on the top 20\% of developers by commit count, who typically contribute a disproportionate share of total project activity.
We validate this choice through sensitivity analysis across 10\%, 20\%, and 30\% thresholds (see \Cref{tab:cascade_sensitivity}).

\section{Results}
\label{sec:results}

In the following, we present the results of our analysis.

\subsection{Burstiness of Commit Activity (RQ1)}

\begin{figure}[htbp]
    \centering
    \includegraphics[width=\linewidth]{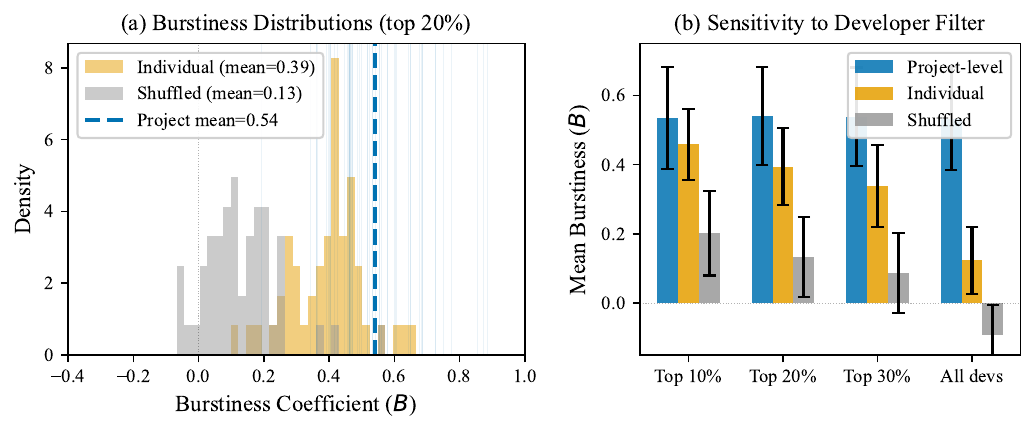}
    \caption{(a) Distribution of per-project mean burstiness coefficients for individual developers (orange) and their shuffled baseline (grey), with the project-level mean shown as a dashed blue line. The clear separation confirms that burstiness is not a product of random timing. (b) Sensitivity to developer selection threshold: project-level burstiness ($B \approx 0.54$) is stable across all filters, while individual-level burstiness naturally decreases as more peripheral developers are included.}
    \label{fig:burstiness_distribution}
\end{figure}

The analysis of 51 projects confirms that commit activity is bursty at both the project and individual levels. At the project level, we find a mean burstiness coefficient $B = 0.54$ (std = 0.14), with all projects showing $B > 0.19$. At the individual developer level, the mean burstiness is $B = 0.39$ (std = 0.11), significantly higher than the shuffled baseline $B_{shuffled} = 0.13$ (std = 0.12). The clear separation between observed and shuffled distributions (Figure~\ref{fig:burstiness_distribution}) confirms that the temporal ordering of commits is non-random, consistent with memory effects where one commit influences the timing of the next.
Project-level burstiness is robust to the developer selection threshold: including all developers ($B=0.53$), the top 30\% ($B=0.54$), or only the top 10\% ($B=0.54$) yields essentially identical results.
This stability is explained by the Pareto principle: the top 10\% of developers already contribute 77\% of commits on average, so the commit stream,  and hence its temporal structure,  is dominated by these core contributors regardless of whether peripheral developers are included.
Individual-level burstiness naturally decreases from $B=0.46$ (top 10\%) to $B=0.12$ (all developers), reflecting that peripheral developers with fewer, irregularly spaced commits exhibit less bursty patterns.
This motivates our investigation of activity cascades as a candidate mechanism underlying these bursty patterns.

\subsection{Cascade Detection Results (RQ2)}

Table~\ref{tab:pareto_cascade_analysis} presents results for cascade detection across 51 repositories, focusing on cascades initiated by the top 20\% most active developers.
Our analysis reveals that activity cascades are a statistically significant phenomenon in a substantial subset of the analyzed OSS projects. 

\paragraph{Validation of Cascade Detection Method}

The validation of our cascade detection method synthetic networks demonstrates good performance (bottom rows of Table~\ref{tab:pareto_cascade_analysis}). 
The random network shows minimal cascade activity (1 observed vs.\ 5.3 expected, $p=1.000$), confirming that our method does not yield false positives.
The cascade network validation shows 249 cascades vs.\ 186.0 expected ($p=0.010$, $d=8.04$), confirming that our method successfully detects planted cascades.

\paragraph{Statistical Significance of Cascades in Empirical Data}

The permutation testing framework reveals statistically significant cascade effects in 28 out of 51 repositories (55\%) when comparing observed cascade counts against the temporal null model. 
Projects demonstrating the strongest cascade effects include large-scale repositories such as `woocommerce` (1741 observed cascades vs. 337 expected under temporal shuffling, p = 0.010, Cohen's d = 26.95), `birt` (2261 vs. 852, p = 0.010, d = 25.84), and `readability` (301 vs. 8, p = 0.010, d = 23.06). 
These results confirm that activity cascades are not merely random fluctuations but represent genuine patterns of social responsiveness in software development.
Sensitivity analysis across developer thresholds and trigger percentiles (\Cref{tab:cascade_sensitivity}) shows that the proportion of projects with significant cascades ranges from 24\% to 82\%, with higher trigger thresholds detecting more cascades as expected. The qualitative finding,  that a substantial subset of projects exhibits significant cascade effects,  holds across all 9 parameter combinations.

\begin{table}[htbp]
\centering
\caption{Cascade sensitivity: number of projects (out of 51) with statistically significant cascades ($p<0.05$) across parameter combinations.}
\label{tab:cascade_sensitivity}
\small
\begin{tabular}{lccc}
\toprule
 & \multicolumn{3}{c}{Trigger Percentile} \\
\cmidrule(lr){2-4}
Top Developers & 10th & 25th & 50th \\
\midrule
Top 10\% & 12 (24\%) & 25 (49\%) & 38 (75\%) \\
Top 20\% & 13 (25\%) & \textbf{28 (55\%)} & 40 (78\%) \\
Top 30\% & 13 (25\%) & 30 (59\%) & 42 (82\%) \\
\bottomrule
\end{tabular}
\end{table}

\paragraph{Effect Sizes}
Effect size varies dramatically across repositories, ranging from small (Cohen's d $<$ 0.5) to extremely large (d $>$ 20). 
The distribution of effect sizes follows a bimodal pattern, with most repositories showing either very small effects (suggesting minimal cascade activity) or very large effects (indicating dominant cascade behavior).
The largest effect sizes are observed in repositories with substantial cascade activity, such as `woocommerce` (d = 26.95), `birt` (d = 25.84), and `readability` (d = 23.06). 
These extreme effect sizes indicate that cascade behavior, when present, represents a dominant mode of collaborative interaction rather than a subtle statistical pattern.

\paragraph{Project Heterogeneity and Context Dependence}

A striking finding is the substantial heterogeneity in cascade susceptibility across different projects.
To investigate what drives this heterogeneity, we classify projects into three robustness categories based on how many of the 9 sensitivity parameter combinations yield significant cascades: \emph{always} significant (10 projects, 9/9), \emph{never} significant (6 projects, 0/9), and \emph{parameter-dependent} (35 projects).
A Kruskal-Wallis test across the three groups reveals that the primary distinguishing factor is project scale: team size ($p=0.042$) and total co-editing events ($p=0.034$) differ significantly, with always-significant projects having a median of 350 developers and 314K events vs.\ 102 developers and 35K events for never-significant projects.
Commit concentration by top developers ($p=0.398$) and per-developer productivity ($p=0.572$) do not differ between groups.
This indicates that cascades are a scale-dependent phenomenon: larger, more densely connected projects provide more opportunities for activity to propagate through co-editing chains, while smaller projects may lack the interaction density necessary for cascades to emerge, regardless of how concentrated their development effort is.
Qualitative inspection of project governance structures reveals no clear pattern: both groups include corporate-backed projects (e.g., react-native/Meta vs.\ balrog/Mozilla) and community-driven projects (e.g., xfdesktop vs.\ thunar-archive-plugin), suggesting that governance alone does not determine cascade susceptibility.

\paragraph{Role of Top Contributors}

The focus on the top 20\% most active developers reveals their crucial role in initiating cascades. 
These developers contribute a disproportionate share of commits, and our analysis shows that cascades initiated by them are statistically significant in most repositories where cascade effects are detected.

\begin{table}[htbp]
\centering
\caption{Cascades initiated by top 20\% most active developers. Top20\% shows number of top developers, Commit\% shows their share of total commits. Empirical data shows observed values. For temporal shuffling, \#Cascades includes mean, empirical p-value and Cohen\'s d. Avg. Depth and Avg. Devs show mean values. P-values from one-sided permutation tests (100 shuffles). Significance: $^{*}p<0.05$, $^{**}p<0.01$.}
\resizebox{\linewidth}{!}{%
\begin{tabular}{l|cc|ccc|ccccc}
\toprule
 & \multicolumn{2}{c|}{Pareto Stats} & \multicolumn{3}{c|}{Empirical Data} & \multicolumn{5}{c}{Temporal Shuffling} \\
\cmidrule(lr){2-3} \cmidrule(lr){4-6} \cmidrule(lr){7-11}
Dataset & Top20\% & Commit\% & \#Cascades & Avg. Depth & Avg. Devs & \multicolumn{3}{c}{\#Cascades} & Avg. Depth & Avg. Devs \\
\cmidrule(lr){7-9}
 & & & & & & Mean & p-val & d &  &  \\
\midrule
\textbf{amp-wp} & \textbf{19} & \textbf{95.5} & \textbf{87} & \textbf{2.00} & \textbf{2.92} & \textbf{26.04} & \textbf{$0.040^{*}$} & \textbf{1.52} & \textbf{2.01} & \textbf{2.70} \\
ansible & 483 & 88.1 & 333 & 2.04 & 2.62 & 301.52 & 0.317 & 0.40 & 2.02 & 2.92 \\
azure-sdk-for-node & 27 & 89.8 & 0 & 0.00 & 0.00 & 2.93 & 1.000 & -0.54 & 0.90 & 1.34 \\
balrog & 13 & 86.5 & 0 & 0.00 & 0.00 & 10.62 & 1.000 & -0.60 & 1.86 & 2.33 \\
binaryninja-api & 15 & 91.0 & 35 & 2.00 & 2.91 & 61.12 & 0.832 & -0.39 & 2.17 & 2.80 \\
\textbf{birt} & \textbf{27} & \textbf{79.7} & \textbf{2261} & \textbf{2.33} & \textbf{2.58} & \textbf{852.19} & \textbf{$0.010^{**}$} & \textbf{25.84} & \textbf{2.16} & \textbf{2.82} \\
\textbf{browser-compat-data} & \textbf{68} & \textbf{94.2} & \textbf{46} & \textbf{2.00} & \textbf{3.00} & \textbf{15.37} & \textbf{$0.010^{**}$} & \textbf{4.60} & \textbf{2.01} & \textbf{2.86} \\
\textbf{calendar} & \textbf{16} & \textbf{93.9} & \textbf{250} & \textbf{2.00} & \textbf{3.00} & \textbf{12.89} & \textbf{$0.010^{**}$} & \textbf{17.76} & \textbf{1.88} & \textbf{2.62} \\
\textbf{cloud-init} & \textbf{38} & \textbf{90.6} & \textbf{478} & \textbf{2.57} & \textbf{2.72} & \textbf{236.55} & \textbf{$0.040^{*}$} & \textbf{1.61} & \textbf{2.10} & \textbf{2.79} \\
\textbf{cosmos-sdk} & \textbf{65} & \textbf{90.0} & \textbf{2676} & \textbf{2.13} & \textbf{2.85} & \textbf{963.17} & \textbf{$0.010^{**}$} & \textbf{12.49} & \textbf{2.08} & \textbf{2.97} \\
\textbf{data} & \textbf{56} & \textbf{91.4} & \textbf{102} & \textbf{2.00} & \textbf{2.53} & \textbf{17.63} & \textbf{$0.020^{*}$} & \textbf{5.82} & \textbf{2.02} & \textbf{2.94} \\
diyHue & 12 & 87.7 & 0 & 0.00 & 0.00 & 0.01 & 1.000 & -0.10 & 0.02 & 0.03 \\
\textbf{droolsjbpm-integration} & \textbf{22} & \textbf{79.6} & \textbf{228} & \textbf{2.28} & \textbf{2.39} & \textbf{70.52} & \textbf{$0.050^{*}$} & \textbf{2.54} & \textbf{2.03} & \textbf{2.87} \\
enzyme & 18 & 88.1 & 0 & 0.00 & 0.00 & 2.78 & 1.000 & -0.56 & 0.74 & 1.10 \\
es.react.dev & 112 & 82.9 & 11 & 2.00 & 3.00 & 40.28 & 0.950 & -0.33 & 2.03 & 2.67 \\
flink & 143 & 89.3 & 3334 & 2.12 & 2.84 & 2189.70 & 0.069 & 0.63 & 2.10 & 3.01 \\
\textbf{fs2} & \textbf{37} & \textbf{92.7} & \textbf{25} & \textbf{2.00} & \textbf{2.96} & \textbf{4.90} & \textbf{$0.010^{**}$} & \textbf{4.85} & \textbf{1.93} & \textbf{2.49} \\
github3.py & 18 & 92.2 & 0 & 0.00 & 0.00 & 0.36 & 1.000 & -0.15 & 0.06 & 0.08 \\
\textbf{gitlabhq} & \textbf{230} & \textbf{95.9} & \textbf{1305} & \textbf{2.03} & \textbf{2.84} & \textbf{607.52} & \textbf{$0.010^{**}$} & \textbf{9.45} & \textbf{2.04} & \textbf{2.98} \\
\textbf{godot} & \textbf{273} & \textbf{93.7} & \textbf{1568} & \textbf{2.12} & \textbf{2.71} & \textbf{675.46} & \textbf{$0.030^{*}$} & \textbf{4.90} & \textbf{2.05} & \textbf{2.96} \\
graphql-dotnet & 12 & 92.9 & 2 & 2.00 & 2.00 & 3.49 & 0.465 & -0.25 & 1.36 & 1.54 \\
\textbf{hadoop} & \textbf{83} & \textbf{82.4} & \textbf{1775} & \textbf{2.09} & \textbf{2.75} & \textbf{581.21} & \textbf{$0.010^{**}$} & \textbf{19.91} & \textbf{2.09} & \textbf{2.95} \\
ignite & 47 & 88.3 & 236 & 2.01 & 2.80 & 299.88 & 0.941 & -1.49 & 2.19 & 2.99 \\
\textbf{infinispan} & \textbf{26} & \textbf{92.7} & \textbf{1886} & \textbf{2.36} & \textbf{3.02} & \textbf{1025.84} & \textbf{$0.040^{*}$} & \textbf{3.00} & \textbf{2.06} & \textbf{2.90} \\
\textbf{ironic} & \textbf{66} & \textbf{90.8} & \textbf{745} & \textbf{2.08} & \textbf{2.90} & \textbf{349.21} & \textbf{$0.010^{**}$} & \textbf{8.08} & \textbf{2.11} & \textbf{3.01} \\
\textbf{keras} & \textbf{103} & \textbf{85.4} & \textbf{544} & \textbf{2.01} & \textbf{2.72} & \textbf{118.09} & \textbf{$0.010^{**}$} & \textbf{7.63} & \textbf{2.04} & \textbf{2.95} \\
kit & 15 & 84.8 & 0 & 0.00 & 0.00 & 0.00 & 1.000 & 0.00 & 0.00 & 0.00 \\
lutris & 39 & 93.1 & 12 & 2.00 & 2.00 & 3.97 & 0.069 & 1.75 & 1.50 & 1.85 \\
\textbf{material-ui} & \textbf{196} & \textbf{91.0} & \textbf{1139} & \textbf{2.04} & \textbf{2.90} & \textbf{320.60} & \textbf{$0.010^{**}$} & \textbf{9.79} & \textbf{2.03} & \textbf{2.89} \\
\textbf{mattermost-mobile} & \textbf{20} & \textbf{77.4} & \textbf{719} & \textbf{2.16} & \textbf{2.59} & \textbf{489.35} & \textbf{$0.010^{**}$} & \textbf{3.00} & \textbf{2.27} & \textbf{2.73} \\
\textbf{mule-integration-tests} & \textbf{18} & \textbf{78.7} & \textbf{47} & \textbf{2.87} & \textbf{2.98} & \textbf{1.21} & \textbf{$0.010^{**}$} & \textbf{11.61} & \textbf{0.86} & \textbf{1.21} \\
nest-simulator & 25 & 84.7 & 18 & 2.61 & 2.00 & 15.42 & 0.347 & 0.24 & 2.03 & 2.82 \\
NewPipe & 103 & 83.9 & 19 & 2.05 & 2.74 & 35.14 & 0.802 & -0.83 & 2.05 & 2.88 \\
\textbf{nodejs-docs-samples} & \textbf{35} & \textbf{85.6} & \textbf{137} & \textbf{2.04} & \textbf{2.39} & \textbf{38.45} & \textbf{$0.010^{**}$} & \textbf{6.44} & \textbf{2.04} & \textbf{2.69} \\
\textbf{omnibus-gitlab} & \textbf{81} & \textbf{92.3} & \textbf{350} & \textbf{2.03} & \textbf{2.76} & \textbf{133.43} & \textbf{$0.010^{**}$} & \textbf{6.14} & \textbf{2.06} & \textbf{2.89} \\
OpenTTD-patches & 30 & 96.9 & 11 & 2.00 & 2.00 & 14.62 & 0.465 & -0.15 & 2.00 & 2.89 \\
pandas & 264 & 90.6 & 978 & 2.02 & 2.71 & 632.94 & 0.099 & 1.06 & 2.06 & 2.96 \\
pay-selfservice & 15 & 77.1 & 55 & 2.00 & 2.73 & 45.63 & 0.248 & 0.56 & 2.03 & 2.88 \\
\textbf{prebid-server} & \textbf{48} & \textbf{69.3} & \textbf{859} & \textbf{2.43} & \textbf{2.42} & \textbf{472.81} & \textbf{$0.020^{*}$} & \textbf{3.53} & \textbf{2.15} & \textbf{2.78} \\
pyomo & 19 & 93.6 & 0 & 0.00 & 0.00 & 6.85 & 1.000 & -1.24 & 1.77 & 2.47 \\
Python & 80 & 66.0 & 3 & 2.00 & 3.00 & 1.40 & 0.198 & 0.82 & 1.14 & 1.54 \\
\textbf{qemu} & \textbf{310} & \textbf{92.3} & \textbf{55} & \textbf{2.00} & \textbf{2.53} & \textbf{32.16} & \textbf{$0.040^{*}$} & \textbf{2.22} & \textbf{2.05} & \textbf{2.86} \\
\textbf{qmk\_firmware} & \textbf{219} & \textbf{82.8} & \textbf{580} & \textbf{2.09} & \textbf{2.99} & \textbf{160.79} & \textbf{$0.010^{**}$} & \textbf{8.07} & \textbf{2.04} & \textbf{2.90} \\
\textbf{react-native} & \textbf{267} & \textbf{88.5} & \textbf{988} & \textbf{2.03} & \textbf{2.76} & \textbf{283.29} & \textbf{$0.010^{**}$} & \textbf{13.69} & \textbf{2.04} & \textbf{2.99} \\
\textbf{readability} & \textbf{6} & \textbf{76.1} & \textbf{301} & \textbf{2.00} & \textbf{2.00} & \textbf{7.95} & \textbf{$0.010^{**}$} & \textbf{23.06} & \textbf{0.60} & \textbf{0.74} \\
\textbf{sonic-buildimage} & \textbf{74} & \textbf{76.4} & \textbf{934} & \textbf{2.39} & \textbf{3.04} & \textbf{359.47} & \textbf{$0.010^{**}$} & \textbf{11.82} & \textbf{2.14} & \textbf{3.00} \\
sphinx & 82 & 94.6 & 2 & 2.00 & 2.00 & 5.90 & 0.663 & -0.37 & 1.72 & 2.26 \\
thunar-archive-plugin & 15 & 62.3 & 0 & 0.00 & 0.00 & 0.77 & 1.000 & -0.43 & 0.48 & 0.71 \\
\textbf{woocommerce} & \textbf{183} & \textbf{94.3} & \textbf{1741} & \textbf{2.07} & \textbf{2.82} & \textbf{336.59} & \textbf{$0.010^{**}$} & \textbf{26.95} & \textbf{2.05} & \textbf{2.95} \\
\textbf{xfdesktop} & \textbf{46} & \textbf{83.1} & \textbf{183} & \textbf{2.00} & \textbf{2.01} & \textbf{74.46} & \textbf{$0.030^{*}$} & \textbf{3.05} & \textbf{2.03} & \textbf{2.79} \\
zero-to-jupyterhub-k8s & 26 & 93.6 & 0 & 0.00 & 0.00 & 0.86 & 1.000 & -0.50 & 0.72 & 0.83 \\ \hline
random network & - & - & 1 & 2.0 & 3.00 & 5.3 & 1.000 & -1.69 & 2.23 & 3.20 \\
\textbf{cascade network} & - & - & \textbf{249} & \textbf{7.4} & \textbf{8.01} & \textbf{186.0} & \textbf{$0.010^{**}$} & \textbf{8.04} & \textbf{6.30} & \textbf{7.00} \\
\bottomrule
\end{tabular}%
}
\label{tab:pareto_cascade_analysis}
\end{table}

\subsection{Interpretation of Results}

The results from our first two research questions have implications for understanding the collective dynamics of collaboration in OSS communities.

\textbf{RQ1: Burstiness of Commit Activity.}
Our analysis confirms that commit activity is bursty, not random. The presence of burstiness at both the project and individual levels indicates memory effects, where one action influences the timing of the next. This motivates the search for candidate mechanisms underlying these bursty patterns.

\textbf{RQ2: Cascades as a Candidate Mechanism.}
Our cascade analysis identifies a social mechanism that is consistent with the burstiness observed in RQ1.
The existence of statistically significant activity cascades in over half of the projects answers RQ2 in the affirmative, indicating an implicit coordination mechanism.
While our observational design cannot establish a causal mechanism, the temporal precedence of trigger events combined with significance against a temporal null model that preserves individual activity levels but destroys social ordering establishes a predictive relationship consistent with Granger causality~\cite{granger1969}.
Rather than relying solely on formal communication, developers appear to coordinate through rapid, responsive actions to code modifications from their peers.

The top contributors that initiate most triggers play a dual role: they are not only the main producers of code but also the most frequent initiators of these rapid response chains. This socio-temporal pattern is a measurable correlate of coordination in decentralized development environments.

\section{Developer Churn Prediction  (RQ3)}\label{sec:churn}

While our primary focus is understanding collective social dynamics emerging from the propagation of activity cascades, we now demonstrate the relevance of our insights for developer churn prediction, i.e. for a given developer at a given time $t$ we seek to forecast whether this developer will become inactive in a future time window $[t, t+\delta]$ for some time $\delta$.
This application serves as a validation of our approach and illustrates how insights from cascade analysis can inform method to address practically relevant issues in software development.

\subsection{Churn Prediction Framework}

We evaluate churn prediction using Logistic Regression, an MLP (multi-layer perceptron without message passing), Graph Neural Networks (GCN, GIN, GAT, GraphSAGE), and three causality-aware DBGNN variants~\cite{qarkaxhija2022dbgnn} that perform message passing on higher-order De Bruijn graphs constructed from time-respecting causal walks in the co-editing network.
Logistic Regression and the MLP operate on node features alone, standard GNNs additionally leverage graph topology, and DBGNN acts as a temporal GNN by capturing higher-order temporal dependencies through message passing on De Bruijn graphs that encode the causal ordering of co-editing events.
We compare all models against a \emph{random} baseline that samples labels according to the training class distribution.
Our experimental design follows a leave-one-out cross-validation approach to evaluate cross-project generalizability of churn prediction.

For each repository in our dataset, we create non-overlapping time windows of 12 months.
Each time window serves as an independent training instance, where we extract network features from the collaboration patterns within that year and predict developer churn occurring at the window's end.
A developer is considered \emph{active} in a window if they edited another developer's code during that period (i.e., they appear as a source in the co-editing network); developers whose code was only edited by others are not labeled, as the co-editing event alone does not constitute an active contribution.
A developer is labeled as \emph{churned} if they do not appear as active in the subsequent window.
The last time window of each repository is excluded from the dataset, as no successor window exists to determine churn labels.
This temporal segmentation approach ensures that our model learns from diverse collaboration dynamics across different project phases and scales.

We adopt a leave-one-out methodology that excludes one target repository entirely from training, using all time windows from the remaining repositories as training instances. 
Specifically, for a dataset of $n$ repositories, we train on $(n-1) \times w_i$ instances, where $w_i$ represents the number of valid time windows for repository $i$, and evaluate on all time windows of the held-out repository. 
This inductive approach provides a robust assessment of cross-project generalizability while maximizing the utilization of temporal data across repositories.
To address class imbalance, we use balanced class weighting and evaluate using Balanced Accuracy, ensuring equal importance to churn detection and retention prediction.
For GNN models, we perform a hyperparameter search over hidden dimensions, layer depth, learning rate, and dropout.
All experiments use five independent runs with different random seeds, and features are standardized prior to training.

Our 12 node features (\cref{tab:features}) capture three complementary aspects of developer behavior:

\emph{Individual activity} features characterize a developer's own engagement: maximum inactivity time (longest gap since last commit), developer age (tenure since first contribution), and number of unique commits.

\emph{Neighbor activity} features, inspired by our cascade analysis, capture the temporal activity patterns of a developer's collaborators. The neighbor inactivity features (minimum, maximum, and mean of collaborators' longest inactivity periods) quantify the social environment of each developer.
If a developer's collaborators exhibit long periods of inactivity, this may signal reduced social engagement and increased churn risk; conversely, active neighbors may encourage continued participation.

\emph{Network position} features describe a developer's structural role in the co-editing graph: in- and out-degree, betweenness and closeness centrality, and the topological distance and connection strength to the project founder.

\begin{table}[htbp]
\centering
\caption{Repository Statistics.}
\resizebox{\linewidth}{!}{%
\small
\label{tab:repo_stats}
\begin{tabular}{lrrrrrc}
\toprule
Repository & Nodes & Edges & Unique Edges & First Co-edit & Last Co-edit & Windows \\
\midrule
amp-wp & 142 & 155,634 & 761 & 2015-10-05 & 2024-11-12 & 9 \\
ansible & 5,990 & 951,694 & 33,182 & 2012-02-05 & 2024-11-08 & 12 \\
azure-sdk-for-node & 192 & 125,576 & 891 & 2012-01-15 & 2023-04-28 & 11 \\
balrog & 107 & 31,971 & 565 & 2011-06-30 & 2024-10-31 & 13 \\
binaryninja-api & 110 & 37,452 & 610 & 2015-07-26 & 2024-11-11 & 9 \\
birt & 163 & 665,077 & 2,614 & 2005-02-04 & 2024-11-10 & 19 \\
browser-compat-data & 1,052 & 162,097 & 4,603 & 2016-03-31 & 2024-11-09 & 8 \\
calendar & 132 & 53,080 & 653 & 2014-04-23 & 2024-11-11 & 10 \\
cloud-init & 414 & 82,923 & 2,763 & 2009-01-14 & 2024-11-07 & 15 \\
cosmos-sdk & 625 & 398,956 & 7,497 & 2017-01-11 & 2024-11-12 & 7 \\
data & 595 & 149,337 & 4,302 & 2011-12-19 & 2024-11-11 & 12 \\
diyHue & 117 & 15,091 & 414 & 2017-06-28 & 2024-10-26 & 7 \\
droolsjbpm-integration & 149 & 95,536 & 1,359 & 2006-12-19 & 2024-08-07 & 17 \\
enzyme & 332 & 10,533 & 916 & 2015-11-16 & 2024-02-14 & 8 \\
es.react.dev & 2,160 & 91,542 & 7,584 & 2013-05-29 & 2024-11-10 & 11 \\
flink & 1,479 & 1,543,591 & 20,478 & 2010-12-17 & 2024-11-09 & 13 \\
fs2 & 296 & 55,853 & 1,245 & 2013-03-07 & 2024-10-24 & 11 \\
github3.py & 179 & 12,809 & 540 & 2012-07-09 & 2024-10-08 & 12 \\
gitlabhq & 2,484 & 1,019,375 & 27,713 & 2011-10-08 & 2024-11-11 & 13 \\
godot & 2,688 & 645,244 & 24,676 & 2014-02-10 & 2024-11-11 & 10 \\
graphql-dotnet & 163 & 44,832 & 743 & 2015-09-10 & 2024-11-02 & 9 \\
hadoop & 669 & 632,323 & 11,157 & 2009-05-26 & 2024-11-11 & 15 \\
ignite & 354 & 649,374 & 7,868 & 2014-02-21 & 2024-11-12 & 10 \\
infinispan & 222 & 475,904 & 2,071 & 2009-03-23 & 2024-11-08 & 15 \\
ironic & 558 & 123,512 & 6,149 & 2012-01-31 & 2024-11-06 & 12 \\
keras & 1,283 & 152,126 & 5,265 & 2015-03-27 & 2024-11-10 & 9 \\
kit & 208 & 10,011 & 508 & 2015-02-14 & 2023-05-30 & 8 \\
lutris & 360 & 52,049 & 1,271 & 2012-05-17 & 2024-11-11 & 12 \\
material-ui & 3,063 & 493,564 & 17,622 & 2014-08-19 & 2024-11-12 & 10 \\
mattermost-mobile & 165 & 48,836 & 687 & 2020-11-24 & 2024-11-11 & 3 \\
mule-integration-tests & 98 & 165,042 & 1,198 & 2005-03-16 & 2024-11-01 & 19 \\
nest-simulator & 138 & 158,726 & 1,694 & 2015-05-12 & 2024-10-18 & 9 \\
NewPipe & 892 & 98,785 & 6,586 & 2015-09-16 & 2024-11-13 & 9 \\
nodejs-docs-samples & 341 & 75,193 & 1,878 & 2013-01-22 & 2024-11-01 & 10 \\
omnibus-gitlab & 748 & 64,246 & 4,148 & 2013-12-03 & 2024-11-08 & 10 \\
OpenTTD-patches & 243 & 722,513 & 2,482 & 2004-08-10 & 2024-11-12 & 20 \\
pandas & 3,345 & 641,259 & 30,543 & 2011-03-18 & 2024-11-12 & 13 \\
pay-selfservice & 76 & 93,964 & 1,088 & 2015-10-08 & 2024-11-07 & 9 \\
prebid-server & 424 & 52,077 & 2,911 & 2017-05-01 & 2024-11-05 & 7 \\
pyomo & 126 & 158,705 & 584 & 2014-10-11 & 2024-11-08 & 10 \\
Python & 1,070 & 28,118 & 4,014 & 2016-07-20 & 2024-11-11 & 8 \\
qemu & 2,320 & 877,431 & 29,733 & 2006-02-04 & 2024-11-06 & 18 \\
qmk\_firmware & 2,151 & 228,813 & 9,446 & 2012-10-18 & 2022-02-12 & 9 \\
react-native & 3,333 & 564,859 & 28,680 & 2015-01-30 & 2024-11-11 & 9 \\
readability & 79 & 9,224 & 303 & 2015-02-09 & 2024-10-17 & 9 \\
sonic-buildimage & 522 & 71,509 & 4,947 & 2016-03-09 & 2024-11-08 & 8 \\
sphinx & 797 & 118,674 & 3,522 & 2007-08-03 & 2024-11-08 & 17 \\
thunar-archive-plugin & 139 & 4,128 & 496 & 2006-03-27 & 2024-10-08 & 18 \\
woocommerce & 1,533 & 725,328 & 12,149 & 2011-08-25 & 2024-11-13 & 13 \\
xfdesktop & 341 & 157,083 & 2,557 & 2003-01-16 & 2024-11-12 & 21 \\
zero-to-jupyterhub-k8s & 234 & 17,554 & 760 & 2016-11-04 & 2024-11-07 & 8 \\
\bottomrule
\end{tabular}
}
\end{table}

\begin{table}[htb]
\centering
\caption{Node features for churn prediction, grouped by category. Importance values are mean absolute logistic regression coefficients (C=10, trained on all projects).}
\resizebox{\linewidth}{!}{%
\scriptsize
\label{tab:features}
\begin{tabular}{|p{3cm}|p{3.5cm}|p{1.2cm}|}
\hline
\textbf{Feature} & \textbf{Description} & \textbf{$|\beta|$} \\
\hline
\multicolumn{3}{|c|}{\textbf{Individual Activity}} \\
\hline
Maximum Inactivity Time & Longest gap since last commit within window & 1.010 \\
\hline
Developer Age & Time elapsed since first contribution & 0.361 \\
\hline
Unique No. Commits & Total number of distinct commits authored & 0.030 \\
\hline
\multicolumn{3}{|c|}{\textbf{Neighbor Activity (trigger-inspired)}} \\
\hline
Neighbors' Max Inactivity & Max of longest inactivity gaps among collaborators & 0.419 \\
\hline
Neighbors' Mean Inactivity & Mean of longest inactivity gaps among collaborators & 0.258 \\
\hline
Neighbors' Min Inactivity & Min of longest inactivity gaps among collaborators & 0.040 \\
\hline
\multicolumn{3}{|c|}{\textbf{Network Position}} \\
\hline
Out-Degree & Developers whose code this developer edited & 0.694 \\
\hline
Distance to First Contributor & Shortest path to project founder & 0.240 \\
\hline
Closeness Centrality & Inverse mean shortest path distance to all others & 0.215 \\
\hline
In-Degree & Developers who edited this developer's code & 0.094 \\
\hline
Strength to First Contributor & Weighted connection to project founder & 0.085 \\
\hline
Betweenness Centrality & Fraction of shortest paths through developer & 0.012 \\
\hline
\end{tabular}
}
\end{table}

\subsection{Experimental Results}

\begin{figure}[htbp]
    \centering
    \includegraphics[width=\linewidth]{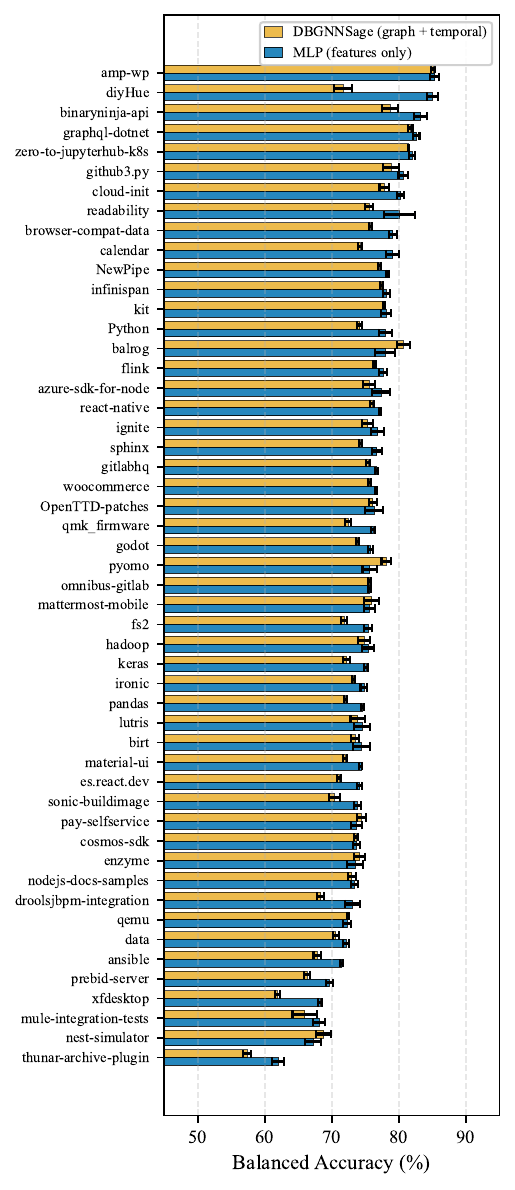}
    \caption{Per-repository balanced accuracy comparing the best MLP (blue) with the best causality-aware GNN, DBGNNSage (orange). Both consume the same 12 node features; DBGNNSage additionally exploits the higher-order co-editing graph. Repositories are sorted by MLP accuracy. The consistent advantage of MLP across the majority of projects indicates that graph structure does not improve prediction beyond what the node features already provide. Error bars show std across 5 runs.}
    \label{fig:churn_per_repo}
\end{figure}

\Cref{tab:churn_results} summarizes the churn prediction performance across all models, with per-repository detail shown in \Cref{fig:churn_per_repo}.
The random baseline reaches 50.2\%, confirming the 50\% floor expected under balanced accuracy.
Logistic Regression reaches 73.7\%, and an MLP without message passing achieves the highest balanced accuracy at 75.8\%.
Among graph-based models, GraphSAGE (74.8\%) is the strongest, while the causality-aware DBGNNSage (73.9\%) outperforms standard GCN (70.4\%) but does not surpass the MLP.
We note that including the validation set in Logistic Regression training yields negligible improvement ($<$0.1\%), confirming that the train/val split used for GNN early stopping does not disadvantage the comparison.

\begin{table}[htbp]
\centering
\caption{Churn prediction results across 51 repositories (leave-one-out cross-validation, balanced accuracy in \%).}
\label{tab:churn_results}
\resizebox{\linewidth}{!}{%
\small
\begin{tabular}{llcccc}
\toprule
Category & Model & Mean $\pm$ Std & Min & Median & Max \\
\midrule
Baseline & Random & $50.2 \pm 1.3$ & 46.1 & 50.1 & 53.0 \\
\midrule
\multirow{2}{*}{Feature-based} & LogReg & $73.7 \pm 4.6$ & 58.5 & 74.2 & 84.4 \\
 & MLP & $75.8 \pm 4.3$ & 62.0 & 75.6 & 85.4 \\
\midrule
\multirow{4}{*}{GNNs} & GCN & $70.4 \pm 5.1$ & 53.8 & 70.7 & 78.7 \\
 & GAT & $73.6 \pm 4.6$ & 55.4 & 73.9 & 83.2 \\
 & GraphSAGE & $74.8 \pm 4.3$ & 61.0 & 75.1 & 83.5 \\
 & GIN & $74.0 \pm 4.8$ & 56.2 & 74.0 & 83.3 \\
\midrule
\multirow{3}{*}{Causality-aware} & DBGNN & $67.5 \pm 6.1$ & 49.8 & 68.0 & 79.8 \\
 & DBGNNSage & $73.9 \pm 4.7$ & 57.4 & 74.2 & 85.1 \\
 & DBGNNGIN & $73.0 \pm 5.0$ & 56.0 & 72.8 & 83.5 \\
\bottomrule
\end{tabular}%
}
\end{table}

\subsection{Feature Importance}

We report feature importances using logistic regression coefficients rather than post-hoc GNN explanation methods, as recent work has shown that GNN explainers produce unfaithful and inconsistent explanations across evaluation dimensions~\cite{agarwal2023evaluating}.
\Cref{tab:features} reports feature importances based on mean absolute logistic regression coefficients.
The strongest predictor is maximum inactivity time ($|\beta|=1.010$), followed by out-degree (0.694) and neighbors' max inactivity (0.419).
While maximum inactivity is by construction correlated with churn, prior work has established that developers who are inactive for more than approximately 295 days have less than a 10\% probability of contributing again~\cite{scholtes2016aristotle, gote2022big}.
Our 12- to 24-month windows are deliberately chosen to capture developers crossing this empirically validated threshold, making maximum inactivity a meaningful predictor rather than a definitional artifact.
The trigger-inspired neighbor features provide a smaller, secondary signal: neighbors' max inactivity ($|\beta|=0.419$) and mean inactivity (0.258) exceed most network position features but fall well below maximum inactivity, indicating that collaborators' disengagement adds only modest predictive value beyond a developer's own behavior.
Traditional productivity metrics like commit count contribute minimally ($|\beta|=0.030$), reinforcing that temporal and social dynamics outweigh raw activity volume in predicting retention.

\subsection{Interpretation of Results}

Our investigation into churn prediction serves as a practical validation for the insights gained from our cascade analysis, directly addressing \textbf{RQ3: Can the presence of activity cascades inform developer churn prediction?} 
The answer is affirmative, but the key insight comes not from the prediction scores themselves, but from understanding which features drive them.

Our models achieve balanced accuracies ranging from 49.8\% (DBGNN, worst repo) to 85.4\% (MLP, best repo) across 51 repositories, with the best model (MLP, 75.8\%) only modestly ahead of the graph-based approaches.
This indicates that the node features carry most of the predictive signal and that message passing over co-editing graphs adds little.
Balanced accuracy is robust to window size (6m: 73.6\%, 12m: 73.7\%, 18m: 74.0\%, 24m: 74.6\%), confirming that the result is not an artifact of the 12-month window choice.
The feature importances show that prediction is dominated by a developer's own maximum inactivity time ($|\beta|=1.010$), with out-degree (0.694) a distant second; the trigger-inspired neighbor features (neighbors' max inactivity 0.419, mean inactivity 0.258) rank above most network-position features but add only a modest secondary signal.
Among GNNs, GraphSAGE (74.8\%) benefits most from graph structure, while GCN (70.4\%) and DBGNN (67.5\%) underperform Logistic Regression, indicating that naive neighborhood aggregation can hurt when co-editing graphs are noisy.
As shown in \Cref{fig:churn_per_repo}, MLP outperforms DBGNNSage in 43 of 51 repositories (mean gap $+1.9$pp) and never loses by more than 2.7pp, reinforcing that propagating signals over the co-editing graph does not, on average, add information beyond each developer's own activity and neighborhood.

We therefore regard churn prediction as a secondary, exploratory contribution rather than a central result.
The dominant predictor, maximum inactivity, is by construction close to the churn outcome, and the trigger-inspired social features add only a modest signal beyond a developer's own (in)activity.
Nonetheless, that the recent inactivity of a developer's collaborators carries any additional predictive value is consistent with our broader finding that developer activity is socially responsive rather than purely individual.

\section{Discussion}
\label{sec:discussion}

Our work provides empirical evidence that temporal contributions of developers in OSS communities exhibit bursty patterns that are consistent with the propagation of activity cascades through co-editing networks.
This has implications for how we understand and model collaborative software development.

\paragraph{Burstiness and Cascades}
The confirmation of bursty activity patterns in RQ1 motivates the search for social mechanisms that could underlie this phenomenon.
Our cascade analysis identifies co-editing cascades as one such candidate mechanism.
The bursty nature of development means that periods of high activity are expected;
our cascade model provides a framework for understanding how these bursts may be initiated and propagated through social interactions.
Establishing a causal link between cascades and burstiness remains an open question for future work.

\paragraph{Activity Cascades as Emergent Coordination}
The results from RQ2 show that activity cascades are statistically significant in over half of the projects studied, suggesting that developers are responsive to each other's work.
In a distributed environment like OSS, where formal coordination can be minimal, such triggered activity may help align efforts and maintain momentum, although our observational data cannot establish the size of this effect.
A cascade could represent a rapid, collaborative bug-fix, a shared effort to complete a feature, or a mentoring interaction where a senior developer's edit prompts a junior developer to quickly follow up.

The heterogeneity of cascade effects across projects is also revealing.
Our analysis shows that project scale is the primary driver: always-significant projects have a median of 350 developers and 314K co-editing events, compared to 102 developers and 35K events for never-significant projects ($p=0.042$).
In contrast, commit concentration and governance structure do not predict cascade presence.
This suggests that cascades emerge from interaction density,  larger teams with more overlapping code changes create more opportunities for activity to propagate,  rather than from specific organizational or cultural factors.
These differences highlight that ``one size fits all'' models of OSS collaboration are inadequate.

\paragraph{Practical Implications for Churn Prediction}
The trigger-inspired neighbor activity features add a small but measurable signal to churn prediction (RQ3), suggesting a potential, if modest, practical use of our framework.
Existing churn prediction approaches typically rely on static developer attributes or structural network features~\cite{lin2017developer, chang2024condygnn}.
Our neighbor activity features add a dynamic, temporal dimension: rather than asking \emph{who} a developer is connected to, they capture \emph{how active} those connections currently are.
A developer whose collaborators are becoming inactive faces somewhat elevated churn risk, which could complement, rather than replace, existing inactivity-based indicators for project managers monitoring team health.
This opens up new avenues for project managers and community health analysts to monitor and foster developer engagement.

\paragraph{Why Features Outperform Graph Structure}
A notable finding is that the MLP (75.8\%), which uses node features without message passing, outperforms all GNNs (\Cref{tab:churn_results}).
This is consistent with recent work showing that well-designed node features can make GNN message passing redundant~\cite{wu2019simplifying, huang2020combining}, and that even untrained message passing layers can match trained GNNs when high-dimensional features are available~\cite{qarkaxhija2025untrained}.
In our setting, two factors explain this result.
First, the neighbor activity features (neighbors' max, mean, and min inactivity) already encode a one-hop summary of each developer's local neighborhood, partially pre-empting what GNN message passing would learn.
Second, all best-performing GNN configurations use only two layers (\Cref{tab:churn_results}), and deeper architectures consistently degrade performance.
This suggests that in co-editing networks, where edges represent heterogeneous interaction types (bug fixes, feature work, refactoring), multi-hop aggregation mixes signals from unrelated collaboration contexts, introducing noise.
Luan et al.~\cite{luan2023when} provide a theoretical basis for this: GNNs improve over feature-based models primarily when graph structure enhances node distinguishability, which requires sufficient label homophily among neighbors.
In co-editing networks, a developer's neighbors include both long-term collaborators and one-time editors, diluting the structural signal.
The practical implication is that for churn prediction in OSS, investing in domain-informed feature engineering yields stronger returns than applying increasingly complex graph architectures.

\paragraph{Limitations and Future Work}
Our study has several limitations. 
First, while our sensitivity analysis (\Cref{tab:cascade_sensitivity}) shows that the qualitative findings hold across parameter choices, the proportion of projects with significant cascades varies substantially (24\%--82\%) depending on the trigger threshold and developer filter, and future work could explore adaptive, per-project thresholds.
Second, we find that cascades are a significant phenomenon in more than half of the projects under the default configuration, which leaves open the question of what mechanisms drive project-level burstiness in projects where we do not find significant cascades.
We hypothesize that this is due to the rather conservative threshold applied in the trigger and cascade detection method.
Third, our pipeline does not explicitly filter bot-generated commits (e.g., dependabot, renovate). In projects with heavy bot activity, automated commits could introduce spurious co-editing edges; however, bots rarely edit the same lines as human developers, limiting their impact on co-editing networks.
More fundamentally, our observational design does not rule out external factors, such as coordinated release cycles, shared deadlines, or continuous-integration events, that could synchronize developer activity and produce apparent triggers without direct social influence; separating such exogenous drivers from genuine social triggering remains an open problem.
Finally, our analysis is based on commit data alone.
Integrating communication data from platforms like Slack, mailing lists, or GitHub issues could provide a richer context for why cascades occur and in which networks they propagate.
Future research could also explore qualitative aspects of cascades. 
Are they primarily about bug-fixing, or feature development? 
Combining our quantitative analysis with qualitative case studies could provide deeper insights into the mechanisms driving this behavior.

\section{Conclusion}
\label{sec:conclusion}

In this work, we present a multi-stage analysis of collective social dynamics in OSS development, moving from a mere description of temporal commit patterns to the underlying social mechanisms of activity cascades.
We first establish that commit activity is inherently bursty, indicating that developer actions are not independent events and suggesting the presence of memory and triggering effects.
This foundational finding motivated our investigation of activity cascades as a social mechanism associated with these dynamics.
We successfully validate the existence of activity cascades in a majority of the 51 OSS projects we analyzed. 
These cascades, where one developer's edit is followed by a chain of rapid responses from collaborators, represent a form of implicit coordination in decentralized development environments.
We also find that these social responsiveness patterns carry modest predictive information for developer churn, though this application is secondary and dominated by individual inactivity.

Our research contributes to the open question how macro-level collective dynamics such as bursty contribution patterns emerge based on micro-level social mechanism.
It provides a new lens through which we can view OSS collaboration, shifting the focus from static network structures to the dynamic, responsive interactions that form the social fabric of these communities. 
By understanding the propagation of activity, we gain deeper insights into project health, developer engagement, and the complex interplay between social and technical contributions in the world of open source.
Methodologically, our churn results also show that node features alone can match graph neural networks on these co-editing networks.

\bibliographystyle{ieeetr}
\bibliography{main}

\end{document}